\documentclass[twoside,a4paper]{article}
\usepackage{graphicx}
\usepackage{fancyhdr}
\usepackage{multicol}
\usepackage{styl-ap}

\begin{document}
\title{Probing the accretion processes in soft X-ray selected polars}
\author{Iris Traulsen\work{1}, Klaus Reinsch\work{2}, Axel D.~Schwope\work{1}
} \workplace{%
Leibniz-Institut f\"ur Astrophysik Potsdam (AIP), An der Sternwarte 16, 14482
Potsdam, Germany
\next
Institut f\"ur Astrophysik, Georg-August-Universit\"at G\"ottingen,
Friedrich-Hund-Platz 1, 37077 G\"ottingen, Germany
}
\mainauthor{itraulsen@aip.de}
\maketitle

\begin{abstract}%
  High-energy data of accreting white dwarfs give access to the regime of the
  primary accretion-induced energy release and the different proposed
  accretion scenarios. We perform XMM-Newton observations of polars selected
  due to their ROSAT hardness ratios close to -1.0 and model the emission
  processes in accretion column and accretion region. Our models consider the
  multi-temperature structure of the emission regions and are mainly
  determined by mass-flow density, magnetic field strength, and white-dwarf
  mass. To describe the full spectral energy distribution from infrared to
  X-rays in a physically consistent way, we include the stellar contributions
  and establish composite models, which will also be of relevance for future
  X-ray missions. We confirm the X-ray soft nature of three polars.
\end{abstract}

\keywords{Cataclysmic variables - Polars - Spectroscopy - Photometry - X-rays -
individual: AI Tri, QS Tel, RS Cae}

\begin{multicols}{2}

\section{Introduction}
\label{traulsen-sec1}

  Accretion onto magnetic white dwarfs involves plasma under extreme physical
  conditions, in particular high temperatures up to millions of Kelvin. X-ray
  observations of the discless AM~Her-type systems (``polars'') provide direct
  insight into the accretion processes and the opportunity to study related
  system properties. Hard X-ray emission ($E>0.5$\,keV) arises from the
  cooling accretion column above the white dwarf and soft ($E<0.5$\,keV) from
  the heated accretion region on the white-dwarf surface. Model calculations
  and recent spectral analyses reveal complex structures of the emission
  regions and a wide range of temperatures and densities.  Several systems are
  found at excesses of soft over hard X-ray flux by factors up to 1\,000,
  which can be interpreted as a sign of inhomogeneous accretion.  Full
  understanding of the accretion processes and the binary system requires
  multi-wavelength data, since the different system components dominate the
  spectral energy distribution (SED) at different wavelengths from infrared up
  to X-rays. In a campaign of dedicated XMM-Newton and optical observations of
  selected AM Her-type systems, we spectrophotometrically study the parameters
  and flux contributions of their components. We concentrate on the conditions
  in the emission regions in the post-shock accretion column and on the heated
  white dwarf, flux and luminosity ratios and their strong dependence on the
  choice of the underlying spectral models. Here, we summarize our work on
  three soft X-ray selected polars and our efforts to establish consistent
  multi-wavelength models.

\section{Observed SEDs}
\label{traulsen-sec2}

  Starting in 2005, we obtained XMM-Newton X-ray and ultraviolet data of
  AI~Tri, QS~Tel, and RS~Cae (obs.\ IDs 0306840901, 0306841001, 0404710401,
  0554740801; Traulsen et al., 2010, 2011, 2014), covering one to five orbital
  cycles per object. On the basis of optical monitoring, the TOO observations
  were triggered during high and intermediate high states of
  accretion. Optical photometry and -- for QS~Tel -- spectroscopy were
  performed (quasi)simultaneously. To construct the full long-term SEDs of the
  objects from infrared to X-ray wavelengths, we use publicly available archival
  data, in particular of the WISE, 2MASS, HST, FUSE, and ROSAT
  archives. Figure~\ref{traulsen-fig1} shows all data, the main system
  components being marked according to their approximate flux maxima in the
  middle panel as follows: \textit{a.} the secondary star in the IR,
  \textit{b.}  the cyclotron emission in the IR to optical, \textit{c.} the
  accretion stream and \textit{d.} the white-dwarf primary in the optical to
  UV, \textit{e.}  the accretion-heated region on the white-dwarf surface in
  the extreme UV to the soft X-ray regime, and \textit{f.}  the post-shock
  accretion column in hard X-rays.  All panels include both high-state and
  low-state data, identifiable by their different flux levels.  The distinct
  soft X-ray / EUV components at high states are clearly visible, compared for
  example to the low-state SED of EF~Eri (Schwope et al.\ 2007) or to the
  serendipitously discovered X-ray hard polar 2XMMp\,J131223.4$+$173659 (Vogel
  et al.\ 2008). At times of high soft X-ray flux, the optical and soft X-ray
  data show pronounced short-term variability (``flickering'').

\begin{myfigure}
\includegraphics[width=\linewidth]{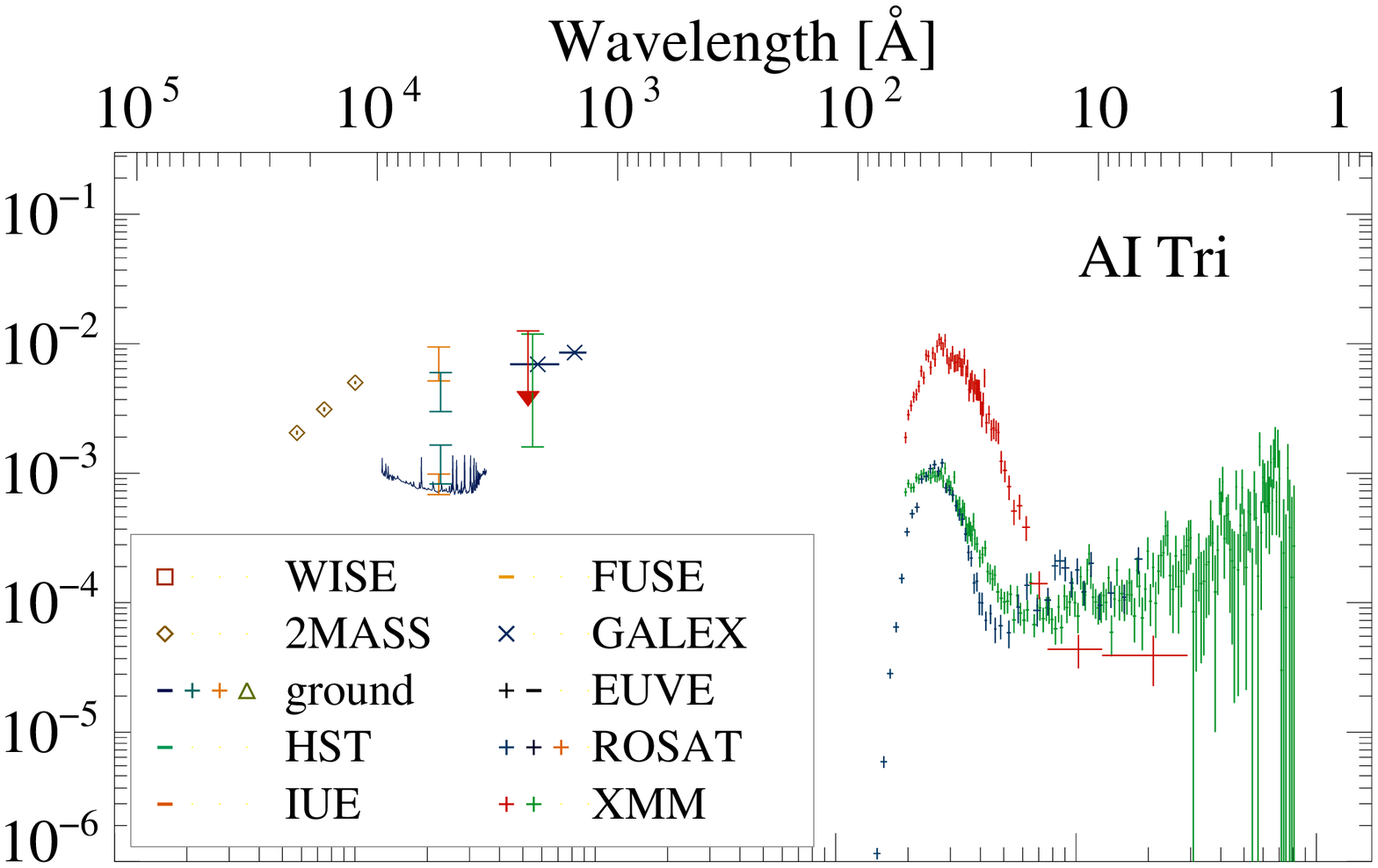}
\includegraphics[width=\linewidth]{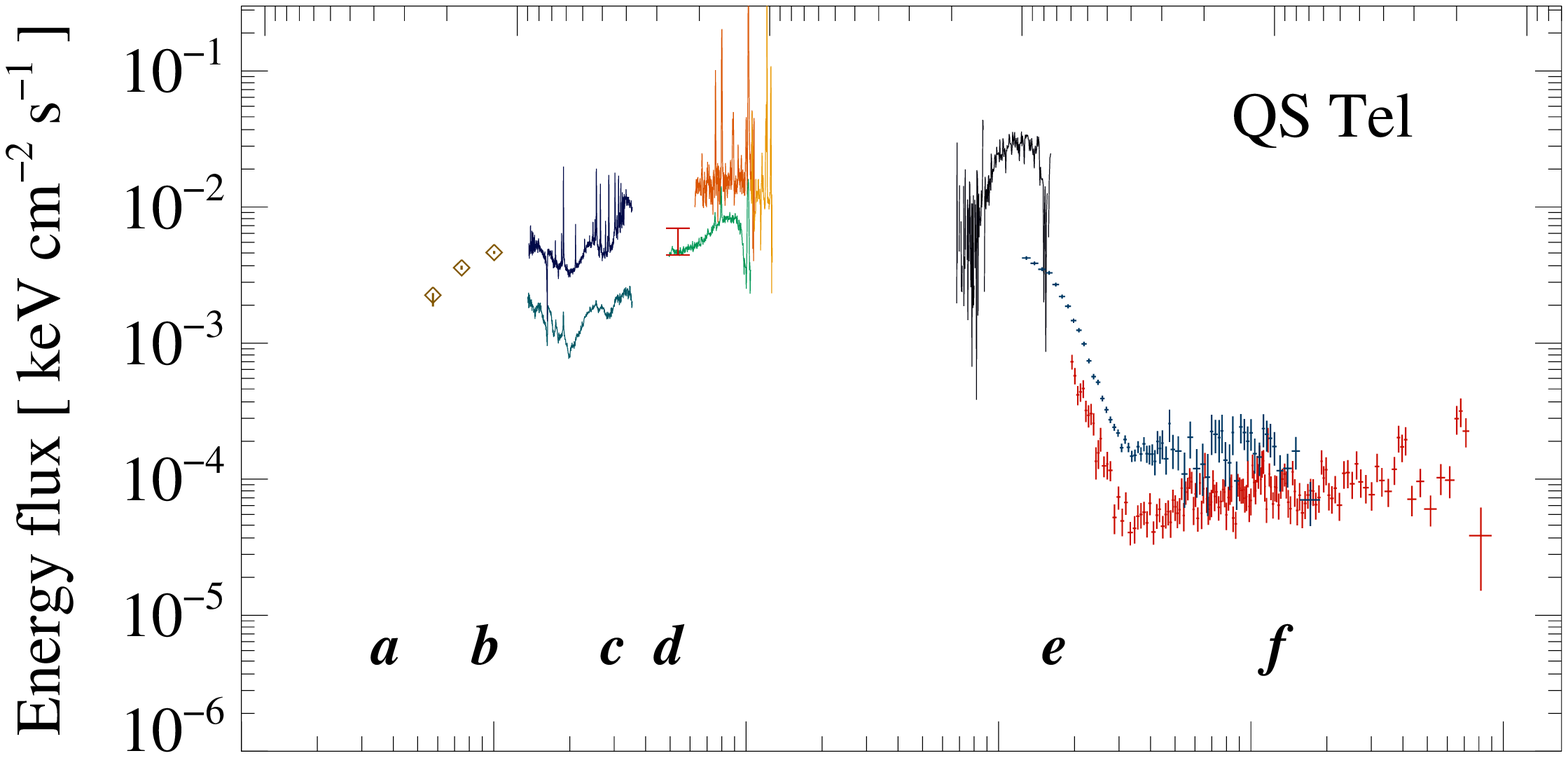}
\includegraphics[width=\linewidth]{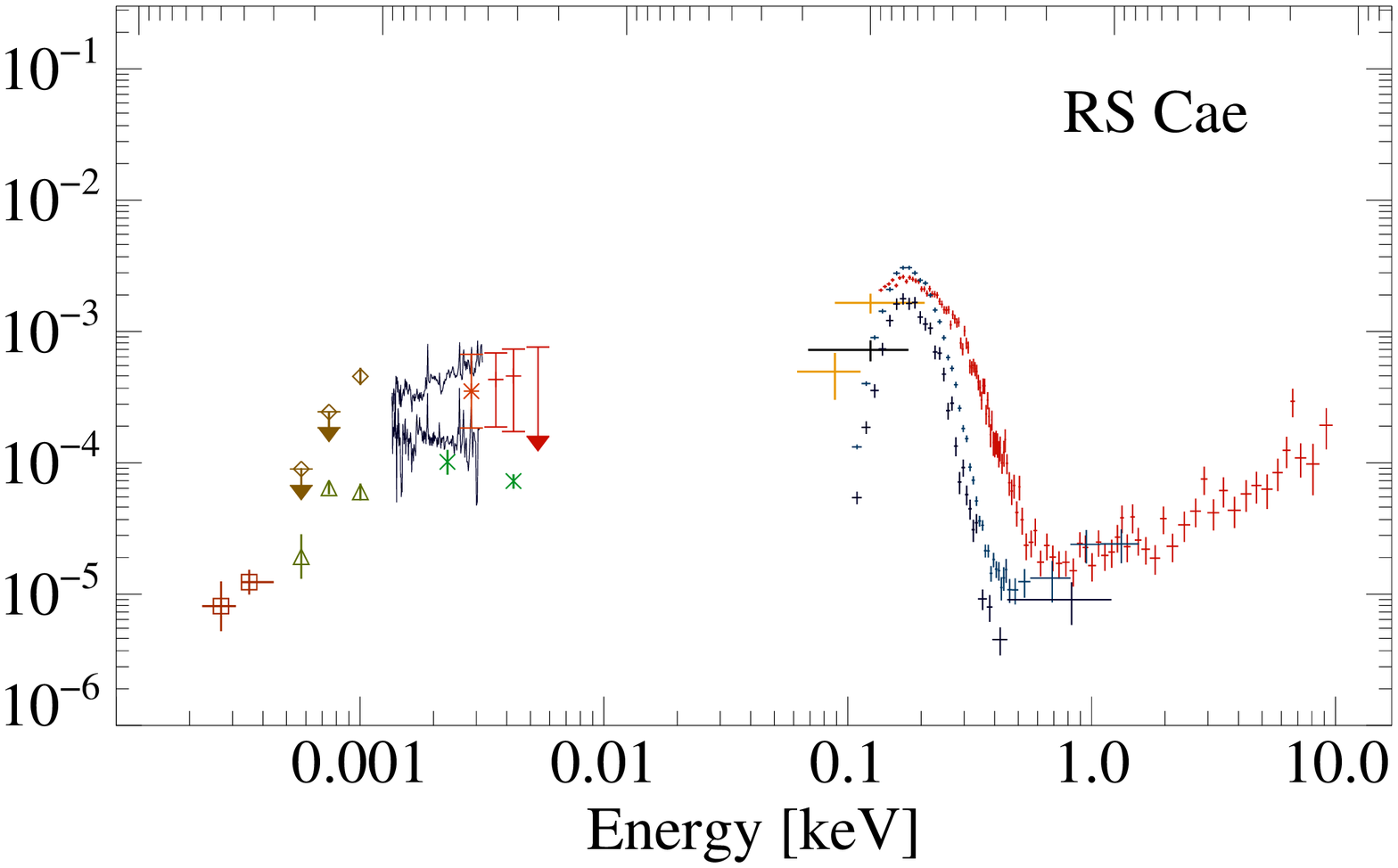}
\caption{Observed SEDs of three X-ray soft polars from IR to X-rays: our
  XMM-Newton and optical plus archival data at different epochs and accretion
  states (cf.\ Traul\-sen et al.\ 2010, 2011, 2014). The X-ray spectra are
  unfolded using the best-fit models.}
\label{traulsen-fig1}
\end{myfigure}

\begin{mytable}
\caption{Parameters of the best fits to the XMM-Newton spectra.  Bolometric
  ROSAT flux ratios at fixed temperatures are calculated from $^{(a)}$Schwarz
  et al.\ 1998, $^{(b)}$Schwope et al.\ 1995, $^{(c)}$Burwitz et al.\ 1996,
  applying corrections of $\kappa_\mathrm{bb}=2.5$, $\kappa_\mathrm{br}=4.8$.}
\label{traulsen-tab1}
\bigskip
\centerline{\begin{tabular}{|@{~}l@{~}|@{~}c@{~}|@{~}c@{~}|@{~}c@{~}|}
\hline
 & AI~Tri & QS~Tel & RS~Cae \\
\hline
black body                            &  multi-T  
                                      &  single-T 
                                      &  single-T  \\
plasma model                          &  multi-T  
                                      &  two-T     
                                      &  single-T  \\
$N_\mathrm{H,ISM}$   \hfill [cm$^{-2}$] & $\sim 10^{20}$ 
                                      & $\sim 10^{20}$
                                      & \parbox{15.5mm}{$2\pm1\times 10^{19}$} \\
$kT_\mathrm{bbody}$  \hfill [eV]       & $44\pm5$
                                      & $20\pm4$
                                      &  $36\pm1$ \\
$kT_\mathrm{plasma}$ \hfill [keV]      & $1\,..\,20$
                                      & $0.2\,..\,4.0$
                                      &  $7\pm3$ \\
$N_\mathrm{H,intr}$  \hfill [cm$^{-2}$] & \parbox{15.5mm}{$3\pm2\times 10^{23}$}
                                      & \parbox{15.5mm}{$10^{22}..10^{23}$}
                                      & \parbox{15.5mm}{$\sim 4\times 10^{23}$} \\
$\log\dot{M}$ [M$_\odot$/yr]           & $-11 .. -9$
                                      &  $\sim -10$
                                      &  $\sim -10$ \\
$F_\mathrm{bb}/F_\mathrm{plasma}$       &  $\sim3 .. 200$
                                      & $\sim10..100$
                                      & $\sim 11$ \\
$F_\mathrm{bb}/F_\mathrm{br,~ROSAT}$    & $\sim170^{(a)}$
                                      & $\sim 80^{(b)}$
                                      & $\sim 90^{(c)}$ \\
\hline
\end{tabular}}
\end{mytable}

\section{Multi-wavelength modeling}
\label{traulsen-sec3}

  To consistently describe the spectral energy distribution and multi-band
  light curves, we synthesize spectral models for one uniform set of system
  parameters and calculate the corresponding light curves. For the different
  system components, we adopt (referring to the labels used in
  Sect.~\ref{traulsen-sec2} and Fig.~\ref{traulsen-fig1}): \textit{a.} a
  PHOENIX stellar atmosphere model of an M star (Hauschildt \& Baron 1999),
  \textit{b.} a cyclotron component of a stratified post-shock accretion
  column (Fischer \& Beuermann 2001), \textit{c.} a simplified
  accretion-stream component of a 3D binary model (Staude et al.\ 2001),
  \textit{d.} a non-LTE white-dwarf atmosphere model (Werner \& Dreizler
  1999), \textit{e.} a single- or multi-temperature black body 
  or hot
  white-dwarf atmosphere, \textit{f.} a single- or multi-temperature plasma
  model of the accretion column.

  The input parameters of the models are determined from observational data,
  where possible, and estimated as typical values for primary white dwarf and
  secondary M star otherwise (e.g.\ Townsley \& Gaen\-si\-cke 2009, Knigge
  2006). In particular, orbital period, inclination, and magnetic field
  strength are available from optical spectroscopy and polarimetry; parameters
  of the accretion-induced emission are fitted to the X-ray spectra
  (cf.\ Sect.~\ref{traulsen-sec4} and Table~\ref{traulsen-tab1}).

  Synthetic cyclotron light curves are derived from the phase-dependent model
  spectra of the accretion-column \textit{(b)} by folding them with the
  Johnson and XMM-Newton OM filter bandpasses, and white-dwarf and
  accretion-stream light curves from model components \textit{c} and
  \textit{d}. By comparing them with observational data, we determine their
  respective phase shifts and intensities. As an example,
  Fig.~\ref{traulsen-fig2} shows the cyclotron spectra calculated for RS~Cae,
  Fig.~\ref{traulsen-fig3} the corresponding synthetic $UBVRI$ light curves.

  Using a consistent set of parameters for all components, we thus establish a
  physically realistic and consistent multi-wavelength model of the whole
  binary system, missing only the unknown spectral contribution of the
  accretion stream. We describe its successful application to the
  multi-wavelength data of RS~Cae in Traulsen et al.\ (2014). The most
  relevant limitation relates to the different observational epochs of the
  high-state data. While we need low-state spectra in the IR and UV to
  identify the secondary and (unheated) primary star, all high-state data
  should be, ideally, observed simultaneously, due to the high variability of
  the accretion processes.

\section{Probing the accretion processes in X-rays}
\label{traulsen-sec4}

  As described above, X-ray data are not the sole, but the main source of
  information on the accretion mechanisms in magnetic CVs. They give us access
  to
\parfillskip=0pt 

\begin{myfigure}
\includegraphics[width=\linewidth]{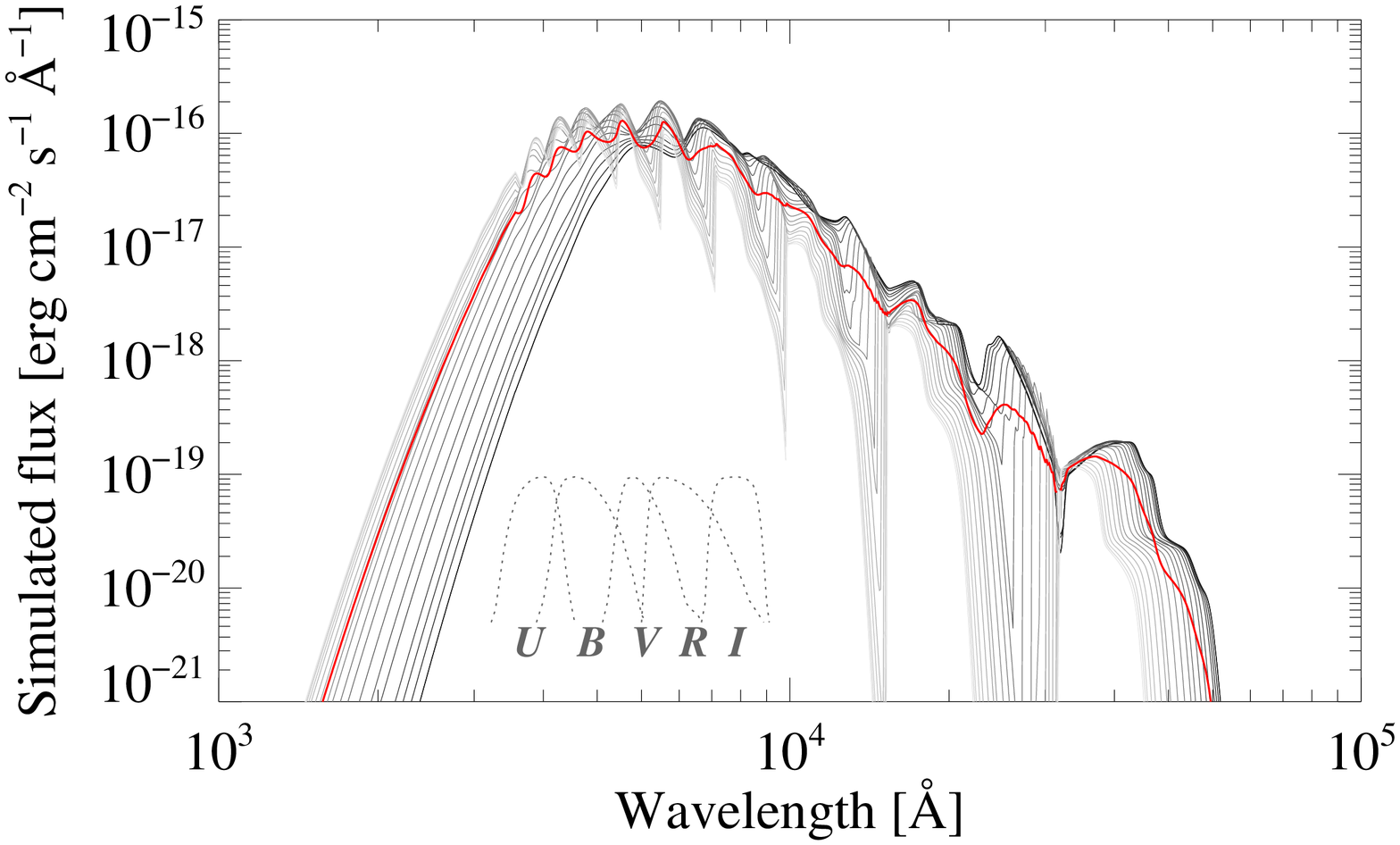}
\caption{Spectral models of cyclotron emission for $B=36\,\mathrm{MG}$,
  $\dot{m}=0.01$, $1.0\,\mathrm{g\,cm^{-2}\,s^{-1}}$, and the estimated
  accretion geometry of RS~Cae: orbital mean (red) and phase-resolved from
  $\varphi_\mathrm{mag}=0.0$ (light) to $0.5$ (dark). Dotted: Johnson filter
  bandpasses, used to derive the light curves in Fig.~\ref{traulsen-fig3}.}
\label{traulsen-fig2}
\end{myfigure}

\noindent 
  characteristic determinants of the systems and their evolution like mass
  accretion rate, component masses, and bolometric fluxes / luminosities,
  which let us distinguish between accretion scenarios like standing or buried
  shocks or inhomogeneous accretion. These objectives, however, are limited by
  the complexity of the puzzle and by the energy resolution and
  signal-to-noise ratio of currently available X-ray data.  Several model
  approaches have been developed, each of them focusing on different aspects,
  as Cropper et al.\ (1999, mass determination and effective spectral
  fitting), Fischer \& Beuermann (2001, column structure, SED coverage). Mass
  and flux determination particularly depend on the underlying spectral models
  (see also Cropper et al.\ 1999).

  The X-ray soft component is usually modeled by an absorbed black body, which
  is on the one hand an appropriate approach for CCD data, not resolving the
  line features. On the other hand, non-LTE processes and the metal-richness
  of the hot photosphere have a non-negligible effect on the spectral
  continuum in the UV to X-rays (cf.\ Rauch 2003 and references therein). The
  relevance of consistent non-LTE modeling, considering line-blanketing by
  heavier elements has been demonstrated for non-accreting hot white dwarfs
  (e.g.\ Traulsen et al.\ 2005). Bolometric fluxes derived from black-body
  fits to XMM-Newton data are lower by factors up to five than from non-LTE
  models, which typically yield lower effective temperatures and hydrogen
  absorptions. Multi-temperature models, designed to reproduce the temperature
  gradient in the heated accretion region, result in increased bolometric
  fluxes by at least $50\,\%$ with respect to single-temperature models. To
  illustrate the differences between the models, we simulate pointed
  observations with the upcoming eROSITA mission (Merloni et al.\ 2012), which
  will have a higher effective area at energies between 0.2 and 2\,keV than
\parfillskip=0pt 

\begin{myfigure}
\includegraphics[width=\linewidth]{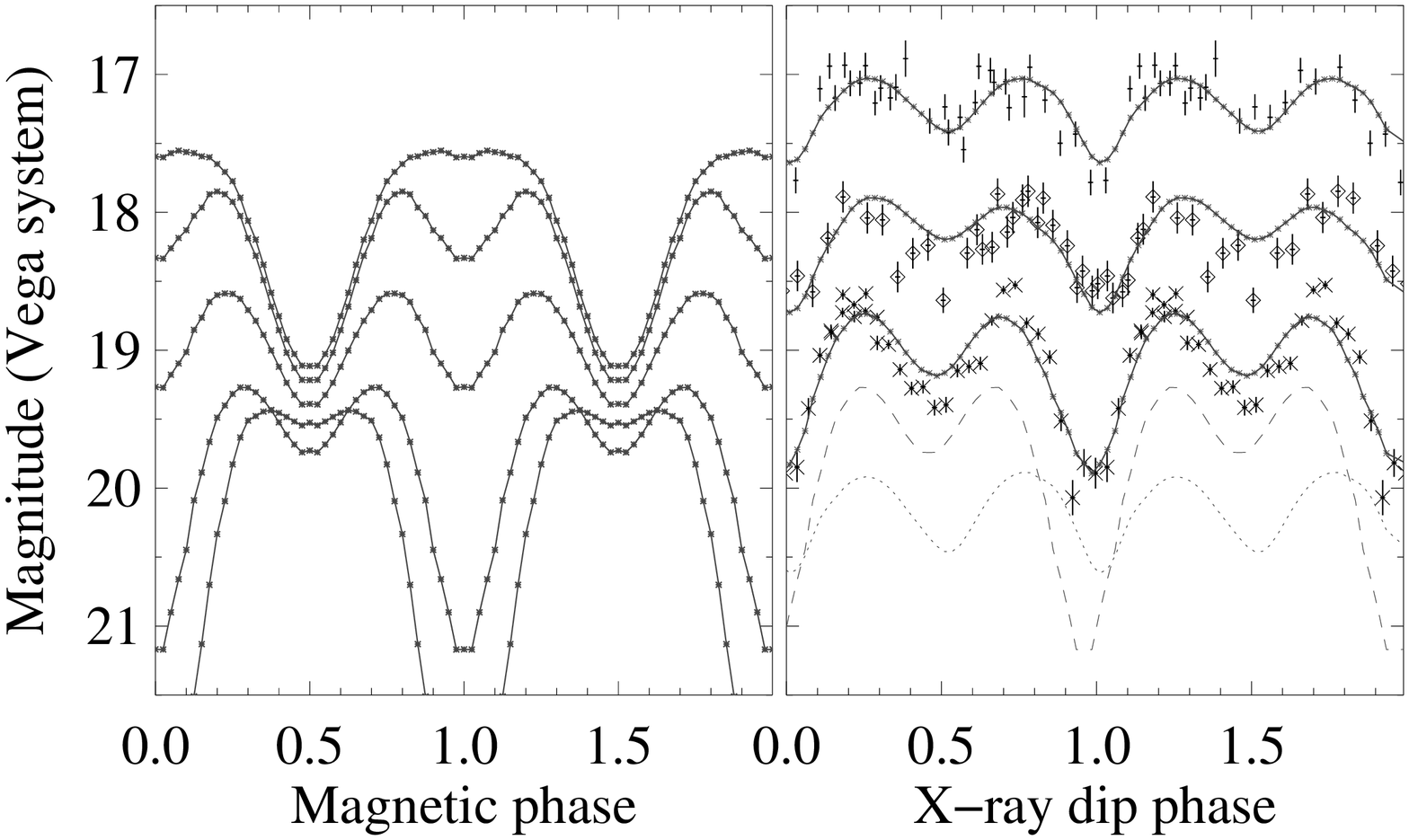}
\caption{Simulated and observed light curves of RS~Cae. \textit{Left:} $UBVRI$
  models (bottom to top), derived from the spectra in
  Fig.~\ref{traulsen-fig2}. \textit{Right:} SMARTS/$B$, XMM-Newton/$U$ and
  UVW1. Dashed: cyclotron emission, dotted: white dwarf and accretion stream.}
\label{traulsen-fig3}
\end{myfigure}

\noindent 
  XMM-Newton and ROSAT. Figure~\ref{traulsen-fig4} shows 50\,ks eROSITA
  spectra based on fits to the XMM-Newton data of AI~Tri (reduced $\chi^2$
  between 0.96 and 1.01).

  For the X-ray hard post-shock spectra, we adopt the radiation-hydrodynamic
  models by Fischer \& Beuermann (2001). They are valid for shallow accretion
  columns, both for the shock scenario dominated by bremsstrahlung cooling and
  the bombardment scenario dominated by cyclotron cooling. To make them
  available for automated spectral fitting of X-ray CCD and grating data, we
  incorporate their temperature and density distributions in \textsc{xspec},
  using the local mass flow densities $\dot{m}$ and magnetic field strengths
  $B$ listed in their paper and white-dwarf masses $M_\mathrm{WD}=0.6$, $0.8$,
  and $1.0\,M_\odot$. We parameterize the distributions for 30 layers of a
  stratified column, and add up 30 \textsc{apec} plasma components to the
  final combined column spectrum. Our models include velocity shifts and
  broadening of the emission lines by stream motion, gravity, orbital motion,
  and the changing viewing angle. Adding a \textsc{pexmon} reflection
  component (Nandra 2007) and multiplying it with the same orbital velocity
  term, we get a comprehensive description of the phase-dependent emission
  induced by the accretion column, in particular of the iron lines between 6.4
  and 6.9\,keV. Figure~\ref{traulsen-fig5} shows composite accretion-column
  models compared to a 50\,ks synthetic spectrum of an (illustrative) polar
  with similar fluxes to AM Her. The models include the same reflection term
  and different column parameters.

  The unabsorbed bolometric fluxes of the column models are typically by about
  $50\,\%$ higher than of the corresponding single-temperature fits. The
  intrinsic absorption and reflection components have a considerable impact on
  the fluxes (cf.\ Cropper et al.\ 1999), which increase by factors up to 15
  compared to pure plasma models. The soft-to-hard flux ratios, thus, may
  significantly vary for the same object and observation, depending on the
  model choice.

\begin{myfigure}
\includegraphics[width=\linewidth]{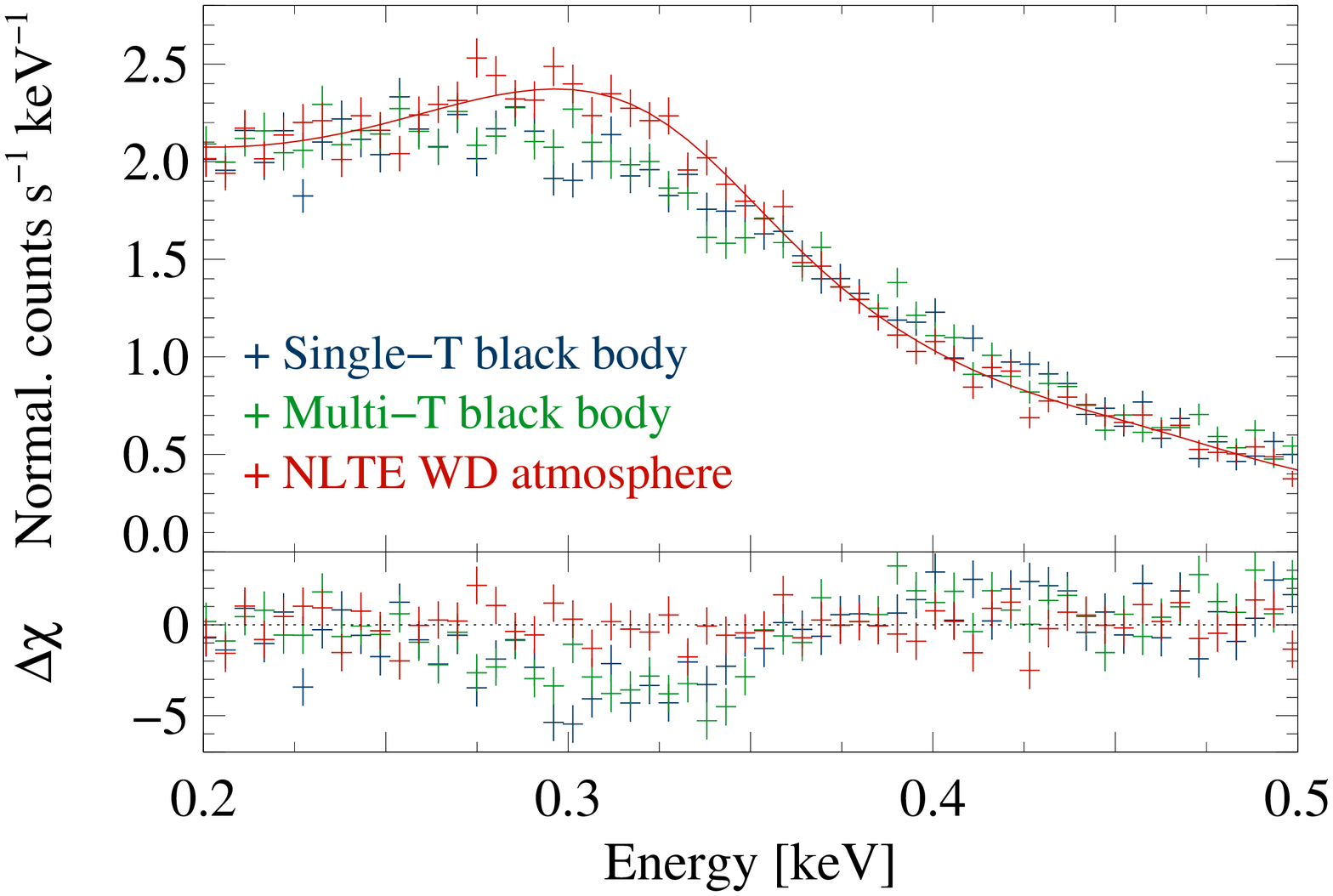}
\caption{Synthetic spectra of a 50\,ks eROSITA pointed observation, based on
  different models for the heated white dwarf in the X-ray soft polar AI~Tri.}
\label{traulsen-fig4}
\end{myfigure}

\begin{myfigure}
\includegraphics[width=\linewidth]{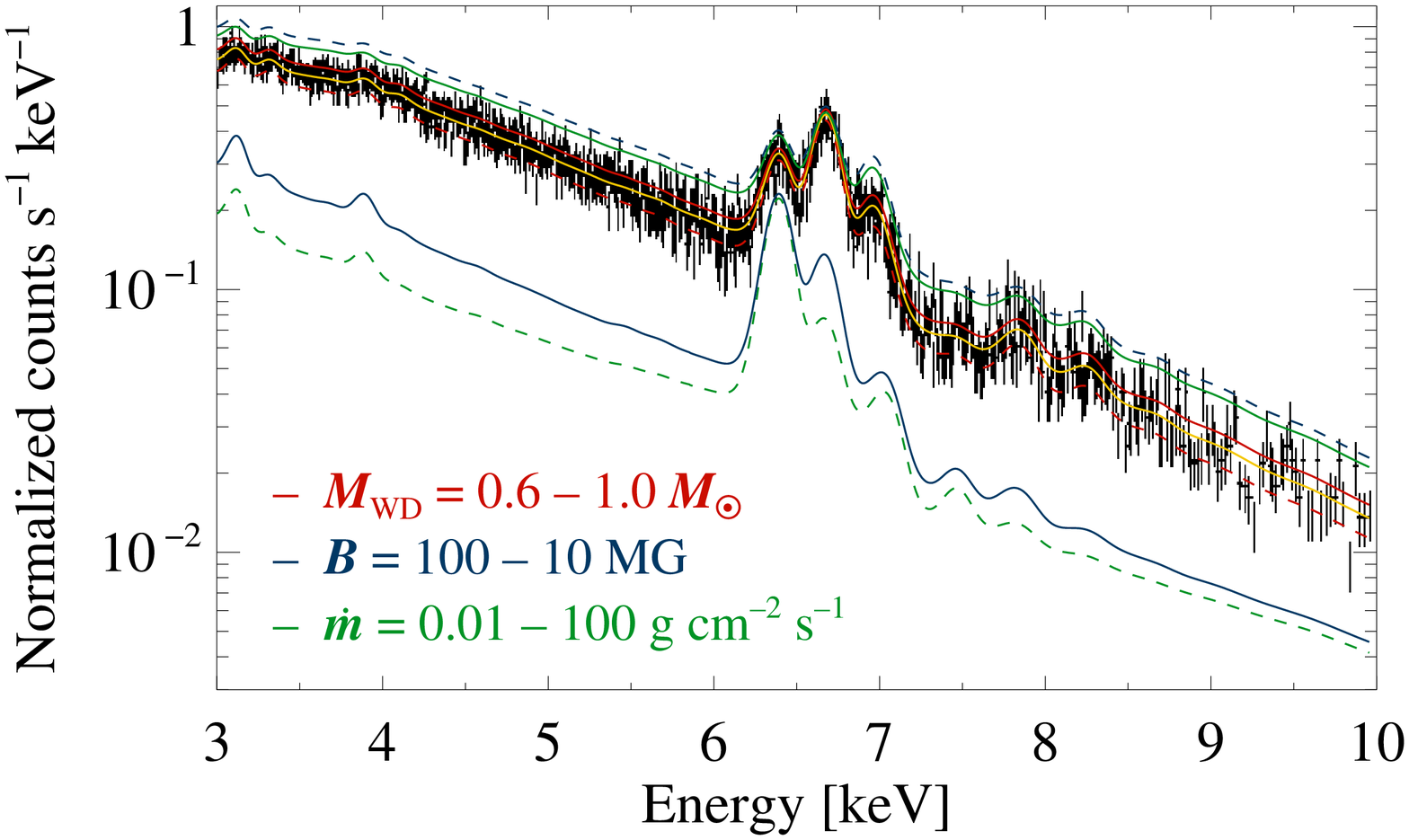}
\caption{Simulated 50\,ks eROSITA observation of a polar at
  $\dot{m}=1.0\,\mathrm{g\,cm}^{-2}\,\mathrm{s}^{-1}$, $B=30\,\mathrm{MG}$,
  $M_\mathrm{WD}=0.8\,M_\odot$, scaled to the XMM-Newton spectrum of AM~Her
  (black: data, yellow: model). Each colored line represents one parameter
  that is varied while the other parameters and the reflection component are
  fixed (dashed: lowest, solid: highest value).}
\label{traulsen-fig5}
\end{myfigure}

\section{Conclusions}
\label{traulsen-sec5}

  Our XMM-Newton observations confirm the soft X-ray excess of the three
  selected polars and indicate inhomogeneous accretion processes. We develop
  composite models including the contributions of the stellar atmospheres and
  the X-ray emitting accretion regions for a physically realistic description
  of the binary system and the accretion processes. Simultaneous
  multi-$\lambda$ observations are relevant for a fully consistent SED
  fitting. The upcoming eROSITA survey will significantly increase the total
  number of known systems and of systems for that reliable soft-to-hard ratios
  can be derived. Survey and pointed observations will enable us to better
  distinguish between different models of the accretion processes, refine
  them, and push our knowledge about the physical properties of the X-ray
  emission regions of polars.

\thanks
  Our research was supported by DLR under grant numbers 50\,OR\,0501,
  50\,OR\,0807, and 50\,OR\,1011.

\bigskip
\bigskip
\noindent {\bf DISCUSSION}

\bigskip
\noindent {\bf CHRISTIAN KNIGGE:} Your data on AI~Tri seems to show that the
hard X-ray flux actually decreases as the accretion rate (and soft X-rays)
increase. What is the interpretation of this?

\bigskip
\noindent {\bf KLAUS REINSCH:} It is related to the response of the accretion
shock height to the changing specific accretion rate. In addition, a higher
fraction of inhomogeneous accretion events can suppress X-ray hard
emission. The first observation of AI~Tri during an extremely soft state does
not cover a full binary orbit.

\end{multicols}

\begin{thebibliography}{99}

\bibitem{} {Burwitz}, V. et al.: 1996, A\&A 305, 507

\bibitem{} {Cropper}, M. et al.: 1999, MNRAS 306, 684

\bibitem{} {Fischer}, A., {Beuermann}, K.: 2001, A\&A 373, 211

\bibitem{} {Hauschildt}, P.~H., {Baron}, E.: 1999, J.\ Comp.\ Appl.\ Math.\
  109, 41

\bibitem{} {Knigge}, C.: 2006, MNRAS 373, 484

\bibitem{} {Merloni}, A. et al.: 2012, arXiv:1209.3114

\bibitem{} {Nandra}, K. et al.: 2007, MNRAS 382, 194

\bibitem{} {Rauch}, T.: 2003, A\&A 403, 709

\bibitem{} {Schwarz}, R. et al.: 1998, A\&A 338, 465

\bibitem{} {Schwope}, A.~D. et al.: 1995, A\&A 293, 764

\bibitem{} {Schwope}, A.~D. et al.: 2007, A\&A 469, 1027

\bibitem{} {Staude}, A., {Schwope}, A.~D., {Schwarz}, R.: 2001, A\&A 374,
  588

\bibitem{} {Townsley}, D.~M., {G{\"a}nsicke}, B.~T.: 2009, A\&A 693, 1007

\bibitem{}{Traulsen}, I. et al.: 2005 in 14th European Workshop on White Dwarfs, D.~{Koester} \&
  S.~{Moehler} (eds.), ASP Conf.\ Ser.\ 334, 325

\bibitem{} {Traulsen}, I. et al.: 2010, A\&A 516, A76
                                                                                
\bibitem{} {Traulsen}, I. et al.: 2011, A\&A 529, A116
                                                                                
\bibitem{} {Traulsen}, I. et al.: 2014, A\&A 562, A42

\bibitem{} {Vogel}, J. et al.: 2008, A\&A 485, 787
                                                                                
\bibitem{} {Werner}, K., {Dreizler}, S.: 1999, J.\ Comp.\ Appl.\ Math.\ 109,
  65


\end{thebibliography}
\end{document}